\documentclass[10pt, twocolumn, twoside]{IEEEtran} 

\usepackage{mathptmx, helvet, courier}
\usepackage[T1]{fontenc}

\usepackage{amsmath,amsfonts,bm,bbm,amssymb,framed}

\usepackage{graphicx, psfrag}
\usepackage[font=small]{caption}  
\usepackage[font=small]{subcaption}

\usepackage{color}
\usepackage{xspace}
\usepackage[english]{babel}
\usepackage[numbers,sort&compress]{natbib}
\bibpunct{[}{]}{,}{n}{}{;}

\usepackage{url}
  
\usepackage{tikz}
\usetikzlibrary{fit}
\usetikzlibrary{calc}
\usetikzlibrary{automata}

\graphicspath{{figures/}}

\newcommand \VAPoRS {\text{VAPoRS}\xspace}
\newcommand \BMA    {\text{BMA}\xspace}

\def \kMAP {k^{\, \text{MAP}}}
\def \kMAPhat {{\hat k}^{\, \text{MAP}}}

\DeclareMathAlphabet\PazoBB{U}{fplmbb}{m}{n}

\newcommand \Kcal {\mathcal{K}}
\newcommand \Lcal {\mathcal{L}}
\newcommand \Jcal {\mathcal{J}}
\newcommand{\dotvar}{\bm{\cdot}}
\newcommand{\vect}[1]{\mathbf{#1}}

\newcommand \un    {\PazoBB{1}}
\newcommand \model {\mathcal{M}}

\newcommand{\distrGauss}{\mathcal{N}}

\newcommand{\distrBern}{\mathcal{B}er}

\newcommand \ak  {\vect{a}_{1:k}}

\newcommand \ok  {\bm{\omega}_{1:k}}
\newcommand \D   {\vect{D}}

\newcommand \y   {\vect{y}}

\newcommand{\dB}{\, \mathrm{dB}}
\newcommand{\SNR}{\mathrm{SNR}}

\newcommand{\E}{\bm{\eta}} 

\newcommand{\m}{\bm{\mu}} 
\newcommand{\s}{\bm{\Sigma}} 

\newcommand{\z}{\vect{z}}
\newcommand{\post}{f}
\newcommand \Nset {\mathbb{N}}
\newcommand \Rset {\mathbb{R}}

\newcommand{\XX}{\bm{\Theta}}  
\newcommand{\THset}{\Theta}    

\newcommand \xxM [1]{\bm{\theta}^{(#1)}}
\newcommand \zzM [1]{\z^{(#1)}}

\newcommand \n {\vect{n}}  
\newcommand \am {\bm{a}_{\mu}} 
\newcommand \tm {\bm{t}_{\mu}} 

\newcommand \VectCompSpec {\bm{\theta}_{1:k}} 
\newcommand \ElemCompSpec [1]{\bm{\theta}_{#1}} 
\newcommand \pair {\bm{\theta}}

\DeclareMathOperator{\argmax}{argmax}
\DeclareMathOperator{\argmin}{argmin}

\renewcommand \th {\text{th}}

\newcounter{Remarkcount}
\newenvironment{remark}
{
\stepcounter{Remarkcount} \medbreak \noindent {\bf Remark} \arabic{Remarkcount}.
}
{
}

\begin{document}

\title{Relabeling and Summarizing Posterior Distributions in Signal
  Decomposition Problems when the Number of Components is Unknown}

\author{Alireza Roodaki, Julien Bect and Gilles Fleury\thanks{Alireza
    Roodaki is with LTCI, CNRS, T{\'e}l{\'e}com ParisTech, Paris,
    France. Email: al.roodaki@gmail.com.}\thanks{Julien Bect and
    Gilles Fleury are with E3S---SUPELEC Systems Sciences, Department
    of Signal Processing and Electronic Systems, SUPELEC,
    Gif-sur-Yvette, France. Email:
    firstname.lastname@supelec.fr}\thanks{The results presented here
    are also part of the Ph.D. thesis of the first
    author~\cite{roodaki:2012:phd}.}}

\maketitle


\begin{abstract} 

  This paper addresses the problems of relabeling and summarizing
  posterior distributions that typically arise, in a Bayesian
  framework, when dealing with signal decomposition problems with an
  unknown number of components. Such posterior distributions are
  defined over union of subspaces of differing dimensionality and can
  be sampled from using modern Monte Carlo techniques, for instance
  the increasingly popular RJ-MCMC method. No generic approach is
  available, however, to summarize the resulting variable-dimensional
  samples and extract from them component-specific parameters.
  
  We propose a novel approach, named \textit{Variable-dimensional
    Approximate Posterior for Relabeling and Summarizing} (\VAPoRS),
  to this problem, which consists in approximating the 
  posterior distribution of interest by a ``simple''---but still
  variable-dimensional---parametric distribution. The distance between
  the two distributions is measured using the Kullback-Leibler
  divergence, and a Stochastic EM-type algorithm, driven by the
  RJ-MCMC sampler, is proposed to estimate the parameters.  Two signal
  decomposition problems are considered, to show the capability of
  \VAPoRS both for relabeling and for summarizing variable dimensional
  posterior distributions: the classical problem of detecting and
  estimating sinusoids in white Gaussian noise on the one hand, and a
  particle counting problem motivated by the Pierre Auger project in
  astrophysics on the other hand.

\end{abstract}


\begin{keywords}
  Bayesian inference; Signal decomposition; Trans-dimensional MCMC;
  Label-switching; Stochastic EM.
\end{keywords}

\section{Introduction}\label{sec:intro}

Nowadays, owing to the advent of Markov Chain Monte Carlo (MCMC)
sampling methods~\cite{metropolis:1953:equation, hastings:1970:monte,
  liu:2001:monte, robert:2004:monte}, Bayesian data analysis is
considered as a conventional approach in machine learning, signal and
image processing, and data mining problems---to name but a
few. Nevertheless, in many applications, practical challenges remain
in the process of extracting, from the generated samples, quantities
of interest to summarize the posterior distribution.


Summarization consists, loosely speaking, in providing a few simple
yet interpretable parameters and/or graphics to the end-user of a
statistical method. For instance, in the case of a scalar parameter
with a unimodal posterior distribution, measures of location and
dispersion (e.g., the empirical mean and the standard deviation, or
the median and the interquartile range) are typically provided in
addition to a graphical summary of the distribution (e.g., a histogram
or a kernel density estimate). In the case of multimodal
distributions, summarization becomes more difficult but can be carried
out using, for instance, the approximation of the posterior by a
Gaussian Mixture Model (GMM)~\cite{west:1993:approx}.  Summarizing or
approximating posterior distributions has also been used in designing
proposal distributions of Metropolis-Hastings (MH) samplers in an
adaptive MCMC framework; see, e.g., \cite{haario:2001:adaptive,
  bai:2011:divcon, bard-2012-adapt}.

This paper addresses the problem of summarizing posterior
distributions in the case of some trans-dimensional problems (i.e.,
``problems in which the number of things that we don't know is one of
the things that we don't know''~\cite{green:1995:reversible,
  green:2003:trans}). More specifically, we concentrate on the problem
of signal decomposition when the number of components is unknown,
which is an important case of trans-dimensional problem. Examples of
such problems include the detection and estimation of sinusoids in
white Gaussian noise~\cite{andrieu:1999:jbm} and the related problem
of estimating directions of arrival in array
processing~\cite{larocque:2002:rjmcmc}, the detection of objects in
images~\cite{rue:1999:bayesian, ortner:2007:building}, and the
detection of physical particles (neutrons, muons, \ldots) using noisy
data from various types of sensors, for instance in
spectroscopy~\cite{andrieu:2002:bayesian} or
astrophysics~\cite{auger:1997:report, albrowd:2004:auger}.


Let $\y\,=\,(y_1,\,y_2,\,\ldots,\,y_N)^t$ be a vector of $N$
observations, where the superscript $t$ stands for vector
transposition. In signal decomposition problems, the model space is a
finite or countable set of models, $\model = \{ \model_k,\; k \in
\Kcal \}$, where $k$ denotes the number of components and $\Kcal
\subset \Nset$ the set of its possible values.  It is assumed here
that, under~$\model_k$, there are $k$ components with vectors of
component-specific parameters $\VectCompSpec = (\ElemCompSpec{1},
\ldots, \ElemCompSpec{k}) \in \THset^k$, where $\THset \subseteq
\mathbb{R}^d$ and $\THset^0 = \{ \varnothing \}$.  One feature that
the problems we are considering have in common is the invariance of
the likelihood $p\left(\y\,|\,k,\,\VectCompSpec\right)$ with respect
to permutations (relabeling) of the components, which is called the
``label-switching'' issue in the literature; see, e.g.,
\cite{richardson:1997:bayesian, stephens:2000:label, jasra:2005:label,
  celeux:2000:comp, fruhwirth:2011:dealing}. We will discuss this
issue further in Section~\ref{sec:label}.

In a Bayesian framework, a joint posterior density
$\post\left(k,\,\VectCompSpec \right) \triangleq
p\left(k,\,\VectCompSpec \,|\, \y\right)$ is obtained through Bayes'
formula for the number~$k$ of components and the vector of
component-specific parameters, after assigning prior distributions on
them:
\begin{equation}\label{eq:post}
  \post\left(k,\,\VectCompSpec \right) %
  \;\propto\;
  p\left(\y\,|\,k,\,\VectCompSpec\right)
  p\left(\VectCompSpec\,|\,k\right)p\left(k\right),
\end{equation}
where $\propto$ indicates proportionality. This density is defined
over a variable-dimensional space~$\XX$, which is a union of subspaces
of differing dimensionality, i.e., $\XX = \cup_{k\ge 0} \{k\} \times
\THset^k$.

The posterior density~\eqref{eq:post} completely describes the
information (and the associated uncertainty) provided by the data~$\y$
about the candidate models and the vector of unknown parameters. Since
it is only known up to a normalizing constant in most cases, Monte
Carlo simulation methods, such as the Reversible Jump MCMC (RJ-MCMC)
sampler~\cite{green:1995:reversible}, have been widely used to
approximate it.

\subsection{The label-switching issue}\label{sec:label}

One of the most challenging issues when attempting at summarizing
posterior distributions, that even occurs in fixed-dimensional
situations, is the label-switching phenomenon (see, e.g.,
\cite{richardson:1997:bayesian, stephens:2000:label, jasra:2005:label,
  celeux:2000:comp, fruhwirth:2011:dealing}), which is caused by the
invariance of both the likelihood and the prior distribution under
permutations of the components. As a consequence, the
component-specific marginal posterior distributions are all equal, and
therefore useless for the purpose of summarizing the information
contained in the posterior distribution about individual components.


The simplest way of dealing with the label-switching issue is to
introduce an Identifiability Constraint (IC), such as sorting the
components with respect to one of their parameters;
see~\cite{richardson:1997:bayesian} for more discussion concerning the
use of ICs in the problem of Bayesian analysis of GMM. However, in
most practical examples, choosing an appropriate IC manually is not
feasible. Many relabeling algorithms have therefore been developed to
``undo'' the label-switching effect automatically, but all of them are
restricted to the case of \emph{fixed}-dimensional posterior
distributions; see~\cite{sperrin:2010:label, yao:2011:label,
  fruhwirth:2011:dealing} for recent advances and references.

In variable-dimensional posterior distributions, there is an
extra uncertainty about the ``presence'' of components, in addition to
their location.  This challenging problem has hindered previous
attempts to undo label-switching in the variable-dimensional scenario,
where, according to~\cite{robert:1997:disc} ``\emph{the meaning of
  individual components is vacuous}''. This argument will be clarified
in the following illustrative example.

\subsection{Illustrative example: joint Bayesian detection and
 estimation of sinusoids in white Gaussian noise }\label{sec:example}

\newcommand \acj   {\ensuremath{a_{c, j}}\xspace}
\newcommand \asj   {\ensuremath{a_{s, j}}\xspace}
\newcommand \ac[1] {\ensuremath{a_{c, #1}}\xspace}
\newcommand \as[1] {\ensuremath{a_{s, #1}}\xspace}

In this example, it is assumed that under $\model_k$, the observed
signal $\y$ is composed of~$k$ sinusoidal components observed in white
Gaussian noise. That is, under~$\model_k$,
\begin{equation*}
  y[i]\;=\; \sum_{j=1}^k\, \left( \ac{j} \cos(\omega_ji)
    \,+\, \as{j} \sin(\omega_ji) \right) \,+\, n[i],
\end{equation*}
where \ac{j} and \as{j} are the cosine and sine amplitudes, 
and $\omega_j$ is the radial frequency of the
$j^\th$ sinusoidal component. Moreover, $n$ is a white
Gaussian noise of variance~$\sigma^2$.

The unknown parameters are the number~$k$ of sinusoidal components,
the vectors $\ElemCompSpec{j} = \left( \ac{j}, \as{j}, \omega_j
\right)$ of component-specific parameters, $1 \le j \le k$, and the
noise variance~$\sigma^2$. Thus, $\THset = \Rset^2 \times (0, \pi)$
and $\XX = \left(\cup_{k\ge 0} \{k\} \times \THset^k \right) \cup
\Rset^+$. We use the hierarchical model, prior distributions, and the
RJ-MCMC sampler proposed in~\cite{andrieu:1999:jbm} for this problem;
the interested reader is thus referred to~\cite{andrieu:1999:jbm,
  green:1995:reversible} for more details\footnote{In fact, the
  ``Birth-or-Death'' moves' acceptance ratio provided in the seminal
  paper~\cite{andrieu:1999:jbm} is
  erroneous. See~\cite[Chapter~1]{roodaki:2012:phd}
  or~\cite{roodaki:2012:note} for justification and true expression of
  the acceptance ratio, which is used in this paper.}.


\begin{figure}  
  \begin{center} 
    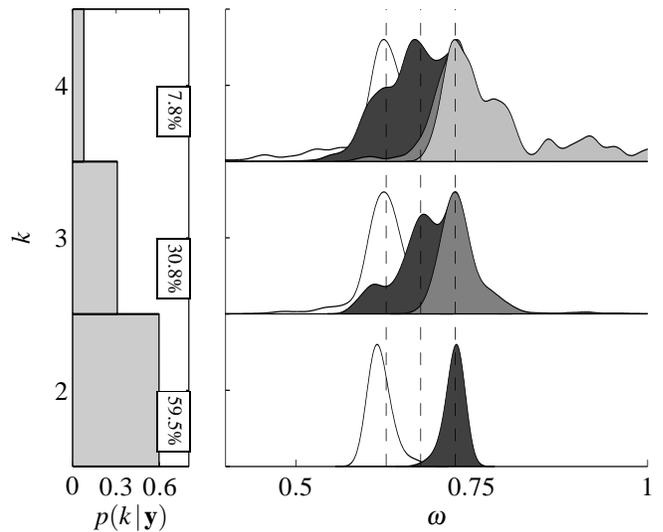
    \caption{Posterior distributions of $k$ (left) and sorted radial
      frequencies, $\ok$, given $k$ (right) from 100\,000 output
      RJ-MCMC samples. The true number of components is three. The
      vertical dashed lines in the right figure locate the true radial
      frequencies.}
    \label{fig:visu}
  \end{center}
\end{figure}

Figure~\ref{fig:visu} represents the posterior distributions of both
the number~$k$ of components and the sorted radial frequencies~$\ok =
\left( \omega_1, \ldots, \omega_k \right)^t$ given~$k$ obtained using
100\,000 samples generated by the RJ-MCMC sampler. Note that, here, we
used sorting to mitigate the effect of label-switching for
visualization. Each row is dedicated to one value of $k$, for $2 \leq
k \leq4$. Observe that other models have negligible posterior
probabilities, since $p(2 \leq k\leq 4 \mid \y) = 0.981$. In the
experiment, the observed signal of length $N = 64$ consists of three
sinusoids with energies $\bm{A}_{1:k} = (20, 6.32, 20)^t$, where $A_j
= \ac{j}^2 + \as{j}^2$, phases $\bm{\phi}_{1:k} = (0, \pi/4,
\pi/3)^t$, where $\phi_j = -\arctan(\as{j}/\ac{j})$, and true radial
frequencies $\ok = (0.63, 0.68, 0.73)^t$. The $\SNR \triangleq \|\D\,
\ak\|^2 \,/\, \left( N\sigma^2 \right)$, where $\ak = \left( \ac{1},\,
  \as{1},\, \ldots,\, \ac{k},\, \as{k} \right)^t$ and $\D$ is the $N
\times 2k$ ``design matrix'' of sines and cosines associated to~$\ok$,
is set to the moderate value of~$7\dB$.

Roughly speaking, two approaches co-exist in the literature for
summarizing variable-dimensional posterior distributions: Bayesian
Model Selection (BMS) and Bayesian Model Averaging (BMA). The BMS
approach ranks models according to their posterior probabilities $p(k
\mid \y)$, selects one model, denoted by~$\kMAP$ here, where MAP
stands for Maximum A Posteriori, and then summarizes the posterior
distribution of the component-specific parameters under the
(fixed-dimensional) selected model.  This is at the price of losing
valuable information provided by the other (discarded) models. For
instance, in the example of Figure~\ref{fig:visu}, all information
about the small---and therefore harder to detect---middle component is
lost by selecting the most \emph{a posteriori} probable
model~$\model_2$. On the other hand, the BMA approach consists in
reporting results that are averaged over all possible models. Although
the BMA approach is suitable for signal reconstruction and prediction
purposes (see, e.g.,~\cite{clyde:2004:model} and references therein),
it is not appropriate for studying component-specific parameters, the
number of which changes in each model\footnote{See, however, the
  intensity plot provided in Section~\ref{sec:result}
  (Figure~\ref{fig:sin-BMA}) as an example of a BMA summary related to
  a component-specific parameter.}. More information concerning these
two approaches can be found in~\cite{clyde:2004:model,
  green:1995:reversible} and references therein.

To the best of our knowledge, no generic method is currently available
that would allow to summarize the information that is so easily read
on Figure~\ref{fig:visu} for this very simple example: namely, that
\emph{there seem to be three sinusoidal components in the observed
  noisy signal, the middle one having a smaller ``probability of
  presence'' than the others}.

\subsection{Outline of the paper}\label{sec:outline}

In this paper, we propose a novel approach, named
\textit{Variable-dimensional Approximate Posterior for Relabeling and
  Summarizing} (\VAPoRS), for relabeling and summarizing posterior
distributions defined over variable-dimensional subspaces that
typically arise in signal decomposition problems when the number of
components is unknown. It consists in approximating the true posterior
distribution with a parametric model (of varying-dimensionality), by
minimization of the Kullback-Leibler (KL) divergence between the two
distributions. A Stochastic Expectation Maximization (SEM)-type
algorithm \cite{celeux:1985:SEM, celeux:1992:SEM, nielsen:2000:SEM},
driven by the output of an RJ-MCMC sampler, is used to estimate the
parameters of the approximate model.

\VAPoRS shares some similarities with the relabeling algorithms
proposed in~\cite{stephens:2000:label, sperrin:2010:label,
  yao:2011:label} to solve the label switching problem, and also with
the EM-type algorithm used in~\cite{bai:2011:divcon} in the context of
adaptive MCMC algorithms (both in a \emph{fixed}-dimensional
setting). The main contribution of this paper is the introduction of
an original variable-dimensional parametric model, which allows to
tackle directly the difficult problem of approximating a distribution
defined over a union of subspaces of differing dimensionality---and
thus provides a first solution to the ``trans-dimensional
label-switching'' problem, so to speak.

Perhaps, the algorithm that we propose can be seen as a realization of
the idea that M. Stephens had in mind when he
stated~\cite[page~94]{stephens:1997:phd}:

``\emph{This raises the question of whether we might be able to obtain
  an alternative view of the [variable-dimensional] posterior by
  combining the results for all different $k$'s, and grouping together
  components which are ``similar'', in that they have similar
  predictive density estimates. However, attempts to do this have
  failed to produce an easily interpretable results.}''

The paper is organized as follows. Section~\ref{sec:VAPORS} introduces
the proposed model and stochastic algorithm for relabeling and
summarizing variable dimensional posterior
distributions. Section~\ref{sec:result} illustrates the performance of
\VAPoRS using two signal decomposition examples, namely, the problem
of joint Bayesian detection and estimation of sinusoids in white
Gaussian noise and the problem of joint Bayesian detection and
estimation of particles in the Auger project (in astrophysics).
Section~\ref{sec:result2} confirms the performances of \VAPoRS using a
Monte Carlo experiment. Finally, Section~\ref{sec:conclusion}
concludes the paper and gives directions for future work.

\section{\VAPoRS} \label{sec:VAPORS}

We assume that the target posterior distribution, defined on the
variable-dimensional space~$\XX = \bigcup_{k \in \Kcal}\, \{k\} \times
\THset^k$, admits a probability density function (pdf) $\post$ with
respect to the $kd$-dimensional Lebesgue measure on each~$\{k\} \times
\THset^k$, $k \in \Kcal$.

Our objective is to approximate the true posterior density~$f$ using a
``simple'' parametric model~$q_{\E}$, where $\E$ is the vector of
parameters defining the model. The pdf~$q_{\E}$ will \emph{also} be
defined on the variable-dimensional space~$\XX$ (i.e., it is not a
fixed-dimensional approximation as in the BMS approach). We assume
that a Monte Carlo sampling method---e.g., an RJ-MCMC
sampler~\cite{green:1995:reversible}---is available to generate~$M$
samples from~$\post$, which we denote by~$\xxM{i} = \bigl( k^{(i)},
\bm{\theta}^{(i)}_{1:k^{(i)}} \bigr)$, for $i \,=\, 1, \ldots, M$.

\subsection{Variable-dimensional parametric model}\label{sec:model}

Instead of trying to describe the proposed parametric family of
densities $\{ q_\eta \}$ directly, let us now adopt a generative point
of view, i.e., let us describe how to sample an $\XX$-valued random
variable~$\pair = ( k, \VectCompSpec )$ from the corresponding
probability distribution. We assume that a positive integer~$L$ is
given, which represents the number of ``components'' present in the
posterior.

First we generate, independently for each of the $L$ components, a
binary indicator variable~$\xi_l \in \{ 0, 1 \}$ drawn from the
Bernoulli distribution $\distrBern(\pi_l)$, where~$\xi_l = 1$
indicates that the corresponding component is \emph{present}
(otherwise, it is \emph{absent}) in~$\pair$. The actual number~$k$ of
components in the generated samples is thus defined as $k =
\sum_{l=1}^L \xi_l$. The parameter $\pi_l \in (0,\,1]$ will be called
the ``probability of presence'' of the~$l^\th$ component.

Second, given the vector of indicator variables~$\bm{\xi} = \left(
  \xi_1,\, \ldots,\, \xi_L \right)$, a $\THset$-valued random vector
is generated for each component that is present (i.e., for each $l$
such that $\xi_l = 1$). This random vector is generated according to
some probability distribution associated to the component, that will
be assumed to be a $\THset$-truncated $d$-dimensional Gaussian
distribution with mean~$\m_l$ and covariance matrix~$\s_l$ in this
paper\footnote{Note that any $d$-dimensional parametric family of
  distributions could be used at this point. As often in the
  literature~\cite{stephens:2000:label, sperrin:2010:label,
    yao:2011:label, bai:2011:divcon}, the Gaussian distribution is
  chosen here as a convenient mean of describing a ``compact'' and
  unimodal $d$-dimensional distribution, nothing more.}. In order to
achieve the required invariance with respect to component relabeling,
the generated vectors are \emph{randomly}\footnote{More precisely, a
  permutation of the $k$~components that are present is drawn
  uniformly in the set of all permutations.} arranged in a vector
$\VectCompSpec = \left( \ElemCompSpec{1}, \ldots, \ElemCompSpec{k}
\right)$.

Contemplating the posterior distributions of the sorted radial
frequencies depicted in the right panel of Figure~\ref{fig:visu},
particularly the plots related to the models with three and four
sinusoidal components, it can be observed that there are ``diffuse
parts'' in the RJ-MCMC output samples resulting in the heavy
asymmetric tails of some components.  It is clear that a model made of
Gaussian components only is not capable of describing these diffuse
samples, at least not in a parsimonious way.  These \emph{abnormal}
observations, with respect to the bulk of the observed data, or,
simply \emph{outliers}, can adversely influence the process of fitting
the approximate posterior to the true posterior distribution of
interest and consequently lead to meaningless parameter estimates.

To overcome this robustness issue, we propose to include in the model
a ``noise-like'' Poisson Point Process (PPP; see, e.g.,
\cite{karr:1991:point}) to account for the presence of outliers in the
observed samples. We assume that the PPP is
homogeneous\footnote{Homogeneity is assumed here for the sake of
  simplicity, but more elaborate (non-homogeneous) models are easily
  accommodated by our approach, if needed.} on~$\THset$, with
intensity $\lambda / |\THset|$. The number of components generated by
the PPP thus follows a Poisson distribution with mean~$\lambda$. To be
consistent with our previous notations, we denote by $\xi_{L+1} \in
\Nset$ this number; note that $\xi_1, \ldots, \xi_L$ still take their
values in~$\{0,1\}$. The (extended) vector~$\bm{\xi}$ thus follows the
probability distribution
\begin{equation}\label{eq:ind_vect}
  p\left(\bm{\xi} \,|\, \bm{\pi}, \lambda\right) \;=\;
  \frac{e^{-\lambda} \cdot \lambda^{\xi_{L+1}}}{\xi_{L+1}!}\,
  \prod_{l=1}^L\pi_l^{\xi_l}(1-\pi_l)^{(1-\xi_l)}.
\end{equation}

\begin{figure}[t]
  \centering \input{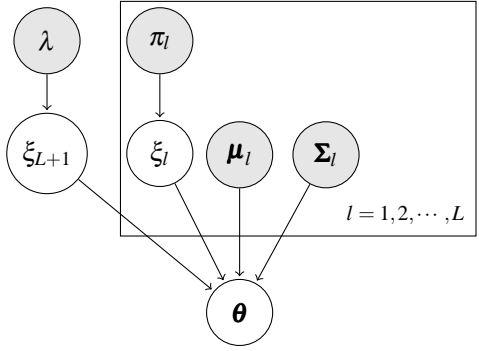}
  \caption{Proposed variable-dimensional parametric model in a
    generative viewpoint.}
  \label{fig:dag_param_model}
\end{figure}

Given~$\bm{\xi}$, $\xi_{L+1}$ random samples are generated uniformly
on~$\THset$ and inserted \emph{randomly} among the samples drawn from
the Gaussian components.  We denote by $q_{\E}$ the pdf of the random
variable~$\pair = ( k, \VectCompSpec )$ that is thus generated, with
$\E = \left( \E_1, \ldots, \E_L, \lambda \right)$ and $\E_l = \left(
  \m_l, \s_l, \pi_l \right)$.  Figure~\ref{fig:dag_param_model}
provides the directed acyclic graph of the model.

\subsection{Distribution of the labeled samples}\label{sec:dist}

A random variable~$\pair = \left(k, \left(\ElemCompSpec{1}, \ldots,
    \ElemCompSpec{k}\right) \right)$ drawn from the density~$q_{\E}$
can be thought of as an ``unlabeled sample'', since the label~$l \in
\Lcal \triangleq \{1,\ldots, L+1\}$ of the component from which
each~$\ElemCompSpec{j}$ ($1 \le j \le k$) originates cannot be
recovered from~$\pair$ itself. Let us now introduce the
(variable-dimensional) \emph{allocation vector}
\begin{equation*}
  \z \;=\ \left(
    k, \left(z_1, \ldots, z_k \right)
  \right)
  \;\in\; \bigcup_{k \in \Kcal} \{k\} \times \Lcal^k,
\end{equation*}
which provides the missing piece of information: $z_j = l$ indicates
that $\ElemCompSpec{j}$ originates from the~$l^\th$ (Gaussian)
component if~$l\leq L$, while $z_j = L+1$ indicates that
$\ElemCompSpec{j}$ originates from the point process component. We
will refer to the pair~$(\pair, \z)$ as a \emph{labeled sample}. In
the following, we will derive its joint distribution~$q_{\E}(\pair,\z)
= q_{\E}(\pair \mid \z)q_{\E}(\z)$.

The distribution of the allocation vector~$\z$ is
\begin{equation}\label{eq:alloc-dist1}
 q_{\E}(\z) \;=\; q_{\E}(\z \mid \bm{\xi})\,q_{\E}(\bm{\xi}),
\end{equation}
where $q_{\E}(\bm{\xi})$ is given in~\eqref{eq:ind_vect}. Note that
$\bm{\xi}$ is a deterministic function of $\z$: $\bm{\xi} = n(\z)$,
with $n_l(\z) = \sum_{j=1}^k \un_{z_j =l}$, for $1 \leq l \leq
L+1$. To compute the first term of~\eqref{eq:alloc-dist1}, remember
that the points generated by the components of the parametric model
are \emph{randomly} arranged in~$\VectCompSpec$. Therefore, for
all~$\bm{\xi} \in \{0,1\}^L \times \Nset$ such that
$\sum_{l=1}^{L+1}\xi_l = k$,
\begin{equation}\label{eq:alloc-dist2}
  q_{\E}(\z \mid \bm{\xi}) \;=\; 
  \frac{\xi_{L+1}!}{k!}\; \un_{\bm{\xi} = n(\z)},
\end{equation}
since two arrangements that differ only by the position of the points
corresponding to the PPP give rise to the same allocation vector.

The conditional distribution $q_{\E}(\pair \mid \z)$ reads
\begin{equation}\label{eq:cond-like}
  q_{\E}(\pair \mid \z) \;=\; \prod_{j=1}^k q_{\E}(\ElemCompSpec{j} \,|\, z_j),
\end{equation}
where
\begin{equation}
  \label{eq:cond-like-j}
  q_{\E}(\ElemCompSpec{j} \,|\, z_j) \;=\;
  \begin{cases}
    \distrGauss\left( \ElemCompSpec{j} \,|\, \m_{z_j},\, \s_{z_j}
    \right)& \text{ if } z_j \,\le\, L,\\
    \frac{1}{|\THset|}& \text{ if } z_j \,=\, L+1.
  \end{cases}
\end{equation}
Therefore, from Equations~\eqref{eq:ind_vect}
to~\eqref{eq:cond-like-j}, we have
\begin{align}\label{eq:comp-like}
  q_{\E}\left(\pair,\, \z\right) & \;=\; \frac{e^{-\lambda}}{k!}\,
  \left(%
    \frac{\lambda}{\left| \THset\right|} \right)^{\xi_{L+1}}\,%
  \prod_{\substack{1 \le j \le k\\ z_j \neq L+1}}\, \distrGauss\left(
    \ElemCompSpec{j} \,|\, \m_{z_j}, \s_{z_j} \right) \nonumber \\
  & \quad \times\; \prod_{l=1}^L\, \pi_l^{\xi_l} \left( 1 \,-\, \pi_l
  \right)^{(1-\xi_l)}\quad \un_{\mathcal{Z}} \left( \z \right),
\end{align} 
where $\left(\xi_1,\, \ldots,\, \xi_{L+1} \right) = n(\z)$ and
$\mathcal{Z}$ is the set of all allocation vectors (i.e., the set of
all $\z \in \cup_{k \in \Kcal} \{k\} \times \Lcal^k$ such that $\xi_l
= n_l(\z) \in \{0, 1\}$, for $1 \leq l \leq L$).

\subsection{Estimating the model parameters}\label{sec:estim}

We propose to fit the parametric distribution~$q_{\E}$ to the
posterior~$f$ of interest by minimizing a divergence measure from~$f$
to~$q_{\E}$. We use the KL divergence as a divergence measure in this
paper, though other divergence measures can be used as
well\footnote{see~\cite[Chapter~2]{roodaki:2012:phd} where another
  divergence measure proposed by~\cite{basu:1998:robust} has been used
  for this problem for robustness reasons.}. 

Denoting the KL divergence from $\post$ to $q_{\E}$ by
$D_{KL}(\post(\pair) \| q_{\E}(\pair))$, we define the criterion to be
minimized as
\begin{equation*} 
  \Jcal(\E) \; \triangleq \; D_{KL}\left(\post(\pair) \,\|\,
    q_{\E}(\pair)\right) \;=\;\int_{\XX} \post\left(\pair\right)
  \,\log\frac{\post(\pair)}{q_{\E}(\pair)} \;\text{d}\pair.
\end{equation*}
Using samples generated by the RJ-MCMC sampler, this criterion can be
approximated as
\begin{equation}\label{eq:kl-crit}
  \Jcal(\E) \; \simeq \; \hat{\Jcal}_M(\E)
  \;=\; -\frac{1}{M}\sum_{i=1}^M \,\log\left(q_{\E}(\xxM{i})\right) + C,
\end{equation}
where $C$ is a constant that does not depend on~$\E$.  One should note
that minimizing $\hat{\Jcal}(\E)$ amounts to choosing
\begin{equation}
  \label{eq:MLE}
  \hat{\E} \;=\; \argmax_{\E}
  \sum_{i=1}^M \log\left(q_{\E}(\xxM{i})\right).
\end{equation}

To estimate the model parameters~$\E \in \mathrm{N}$, one of the
extensively used algorithms for Maximum Likelihood (ML) parameter
estimation in latent variable models is the EM algorithm proposed
by~\cite{dempster:1977:EM}. However, it turns out that the EM
algorithm, which has been used in similar
works~\cite{stephens:2000:label, sperrin:2010:label, bai:2011:divcon},
is not appropriate for solving this problem, as computing the
expectation in the E-step is intricate. More explicitly, in our
problem the computational burden of the summation in the E-step over
the set of all possible allocation vectors~$\z$ increases very rapidly
with both $L$ and $k$.  In fact, even for moderate values of~$L$
and~$k$, say, $L=15$ and $k = 10$, the summation is far too expensive
to compute as it involves $\sum_{m=0}^k \frac{L!}{(L-k+m)!}  \approx
1.3\, 10^{10}$ terms.

In this paper, we propose to use the SEM
algorithm~\cite{celeux:1985:SEM,celeux:1992:SEM, nielsen:2000:SEM}, a
variation of the EM algorithm in which the E-step is substituted with
stochastic simulation of the latent variables from their conditional
posterior distributions given the previous estimates of the unknown
parameters. In other words, at the iteration $r + 1$ of the SEM
algorithm, denoting the estimated parameters at iteration~$r$
by~$\hat{\E}^{(r)}$, for $i\,=\,1,\,\ldots,\,M$, the allocation
vectors $\zzM{i}$ are drawn from
$q_{\hat{\E}^{(r)}}(\bm{\cdot}\,|\,\xxM{i})$. This step is called the
Stochastic (S)-step. Then, these random samples are used to construct
the so-called pseudo-completed log-likelihood.
 
\begin{figure}[t]
  \begin{framed} 
    At the $(r+1)^\th$ iteration, do:\\
    \setlength{\tabcolsep}{3pt}
    \begin{tabular}{ll}
      \\
      (S-step)
      & Draw allocation vectors $\zzM{i,r+1}$, $1 \le i \le M$,\\
      & using an IMH step with target 
      $q_{\hat{\E}^{(r)}}(\dotvar \,|\, \xxM{i})$.\\[10pt]
      (E-step)
      & Construct the pseudo-completed log-likelihood\\[3pt]
      & $\quad 
      \widehat{\Jcal}_M(\E) \;=\; -\sum_{i=1}^{M}\, \log \bigl(
      q_{\E}(\xxM{i},\zzM{i,r+1}) \bigr).$\\[10pt]
      (M-step)
      & Estimate $\hat{\E}^{(r+1)}$ such that\\
      & $\quad 
      \hat{\E}^{(r+1)}\;=\;\argmin_{\E}\,\hat{\Jcal}_M(\E).$
    \end{tabular}
  \end{framed}
  \caption{Proposed SEM-type algorithm}
  \label{fig: SEM}
\end{figure}

Exact sampling from $ q_{\hat{\E}^{(r)}}(\dotvar \mid \xxM{i})$, as
required by the S-step of the SEM-type algorithm, is unfortunately not
feasible---not even using the accept-reject algorithm, due to the
heavily combinatorial expression of the normalizing
constant~$q_{\hat{\E}^{(r)}}(\xxM{i})$. Instead, since
\begin{equation*}
  q_{\hat{\E}^{(r)}}(\zzM{i} \mid \xxM{i}) \;\propto\;
  q_{\hat{\E}^{(r)}}(\xxM{i},\, \zzM{i} )
\end{equation*}
can be computed up to a normalizing constant, we choose to use an
Independent Metropolis-Hasting (IMH) step with $
q_{\hat{\E}^{(r)}}(\zzM{i} \mid \xxM{i})$ as its stationary
distribution; see~\cite[Algorithm~2.2]{roodaki:2012:phd} for more
details. The proposed SEM-type algorithm is summarized in
Figure~\ref{fig: SEM}.

\begin{remark}
  It would also be possible to assign prior distributions over the
  unknown parameters $\E$ and study their posterior distributions (for
  example, using an MCMC sampler with the latent variable~$\z$ added
  to the state of the chain, in the spirit of the ``data
  augmentation'' algorithm \cite{tanner:1987:data}). This would,
  however, leave the label-switching issue unsolved (because of the
  invariance of~$q_{\E}$ to permutations of its components).
\end{remark}

\begin{remark}
  Convergence results of the SEM algorithm in the general form are
  provided by~\cite{nielsen:2000:SEM} and, in the particular example
  of mixture analysis problems, by
  \cite{diebolt:1993:SEM}. Unfortunately, the assumptions in
  \cite{diebolt:1993:SEM, nielsen:2000:SEM} do not hold in the problem
  we are dealing with as, 1) the observed samples~$\xxM{i}$ are
  correlated, owing to the fact that they are generated from the true
  posterior distribution using some MCMC methods, e.g., the RJ-MCMC
  sampler; 2) an I-MH sampler is used to draw~$\zzM{i}$ from the
  conditional posterior distribution. Nevertheless, empirical evidence
  of the ``good'' convergence properties of the SEM-type algorithm we
  proposed will be provided in the next two sections.
\end{remark}

\subsection{Robustified algorithm}\label{sec:robust}

Preliminary experiments with the SEM-type algorithm described in
Figure~\ref{fig: SEM} were not satisfactory, because the sample mean
and (co)variance estimates in the M-step, obtained from minimizing the
KL divergence from the posterior distribution~$f$ to the parametric
model~$q_{\E}$, still suffer from sensitivity to the outliers in the
observed samples, even after including the Poisson point process
component. As a workaround, we propose to use robust
estimates~\cite{huber:2009:robust} of the means and (co)variances of
Gaussian distributions instead of the empirical means and
(co)variances in the M-step. For example, in the case of univariate
Gaussian distributions, one can use the median and the interquartile
range as robust estimators of the mean and variance,
respectively. See~\cite[Section~2.5]{roodaki:2012:phd} for more
discussion of the robustness issue, including an alternative solution
using the ``robust divergence'' of~\cite{basu:1998:robust}.

\begin{remark}
  Similar robustness concerns are widespread in the clustering
  literature; see, e.g., \cite{dave:1997:robust} and references
  therein.
\end{remark}

\section{Illustrative examples} \label{sec:result}

In this section, we will investigate the capability of \VAPoRS for
summarizing variable-dimensional posterior distributions using two
signal decomposition examples; 1) joint Bayesian detection and
estimation of sinusoids in white Gaussian
noise~\cite{andrieu:1999:jbm} and 2) joint Bayesian detection and
estimation of astrophysical particles in the Auger
project~\cite{auger:1997:report, albrowd:2004:auger};
see~\cite[Chapters~3 and~4]{roodaki:2012:phd} for more results and
discussion.  We emphasize again that the output of the
trans-dimensional Monte Carlo sampler, e.g, the. RJ-MCMC sampler in
this paper, is considered as the observed data for \VAPoRS.

\subsection{Joint Bayesian detection and estimation of sinusoids in
  white Gaussian noise}\label{sec:result-sin}

Let us consider the problem of detection and estimation of sinusoidal
components introduced in Section~\ref{sec:example} where the unknown
parameters are the number~$k$ of components, the component-specific
parameters $\left( \ac{j}, \as{j}, \omega_j \right)$, $1 \le j \le k$, 
and the noise variance~$\sigma^2$. Since the amplitudes
and the noise variance can be analytically integrated out,  we focus
on summarizing the joint posterior distribution~$p(k, \,\ok \mid \y)$ of the
form illustrated in Figure~\ref{fig:visu}. Therefore, we assume that
the proposed parametric model introduced in Section~\ref{sec:model}
consists of univariate Gaussian components, with means~$\mu_l$,
variances~$s^2_l$, and probabilities of presence~$\pi_l$, $1 \leq l
\leq L$, to be estimated. Moreover, the space of component-specific
parameters is $\THset = (0, \pi) \subset \Rset$.


Before launching \VAPoRS, we need first to initialize the parametric
model.  It is natural to deduce the number~$L$ of Gaussian components
from the posterior distribution of~$k$. Here, we set it to the
$90^\th$ percentile of~$p(k\mid \y)$ to keep all the probable models
in the play. To initialize the Gaussian components' parameters, i.e.,
$\mu_l$ and $s^2_l$, $1 \leq l \leq L$, we used the robust estimates
of the means and variances of the marginal posterior distributions of
the sorted radial frequencies given~$k\,=\,L$. Finally, we set $\pi_l
= 0.9$, for $1 \leq l \leq L$, and $\lambda = 0.1$.

\begin{figure}[t!]
  \centering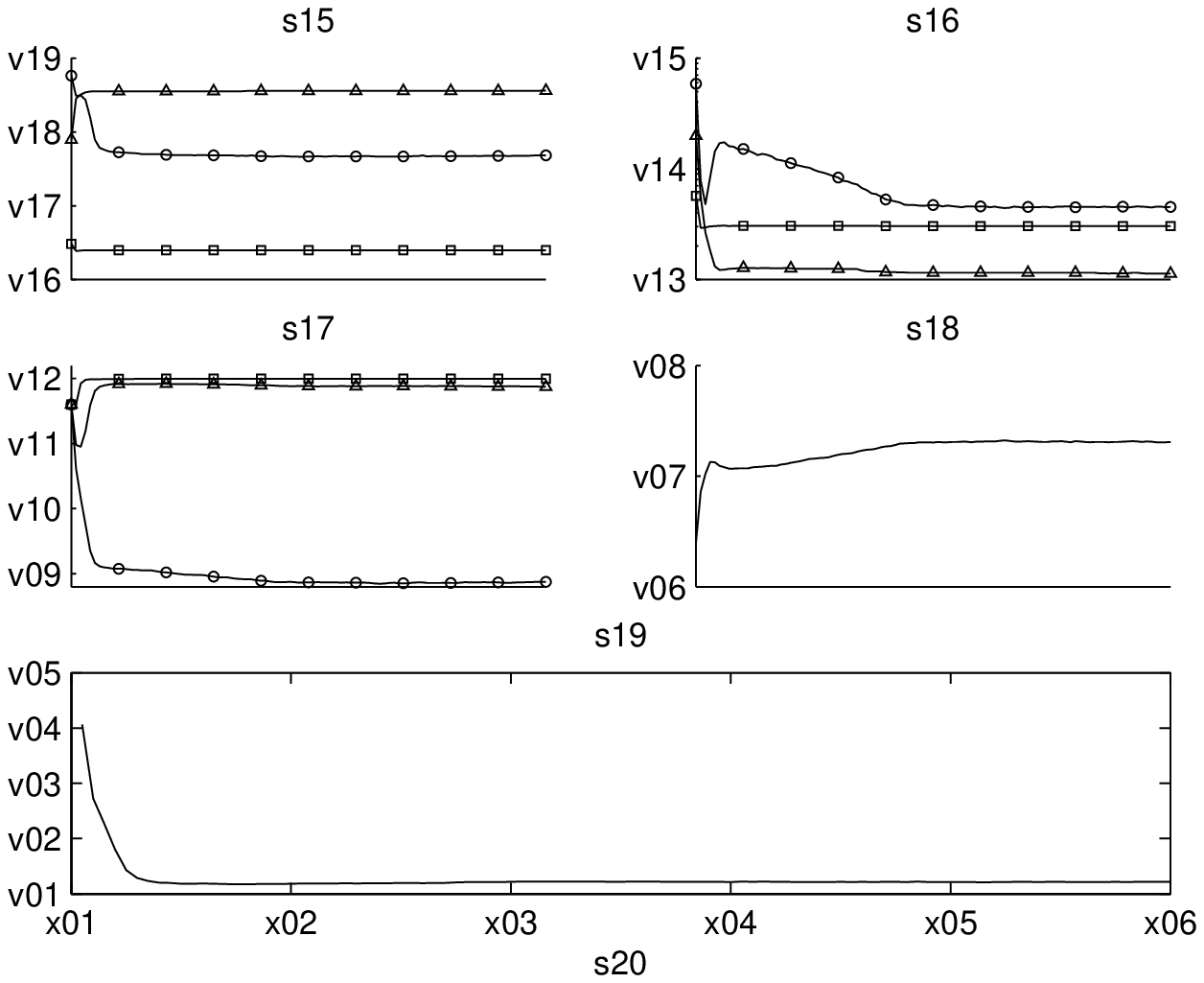
  \caption{Evolution of the model parameters along with the
    criterion~$\hat{\Jcal}_M$ defined in~\eqref{eq:kl-crit} using 100
    iterations of \VAPoRS with $L\;=\;3$ on the RJ-MCMC output samples
    shown in Figure~\ref{fig:visu}.}
  \label{fig:sin-param-evol}
\end{figure}

We ran the ``robustified'' stochastic algorithm introduced in
Section~\ref{sec:VAPORS} on the specific example shown in
Figure~\ref{fig:visu}, for 100 iterations, with $L = 3$ Gaussian
components (note that the posterior probability of $\{ k \le 3 \}$ is
approximately 90.3\%).  To assess the convergence of \VAPoRS,
Figure~\ref{fig:sin-param-evol} illustrates the evolution of the model
parameters~$\E$ together with the criterion~$\Jcal$. Two substantial
facts showing the convergence of \VAPoRS can be deduced from this
figure: first, the decreasing behavior of the
criterion~$\hat{\Jcal}_M$, which is almost constant after the
$10^\th$ iteration; second, the convergence of the parameters of the
parametric model, particularly the means $\mu_l$ and probabilities of
presence $\pi_l$, $1 \leq l \leq L$, even though we used a naive
initialization procedure. Indeed after the $40^\th$ iteration there
is no significant move in the parameter estimates.

\begin{figure}[t!]
  \centering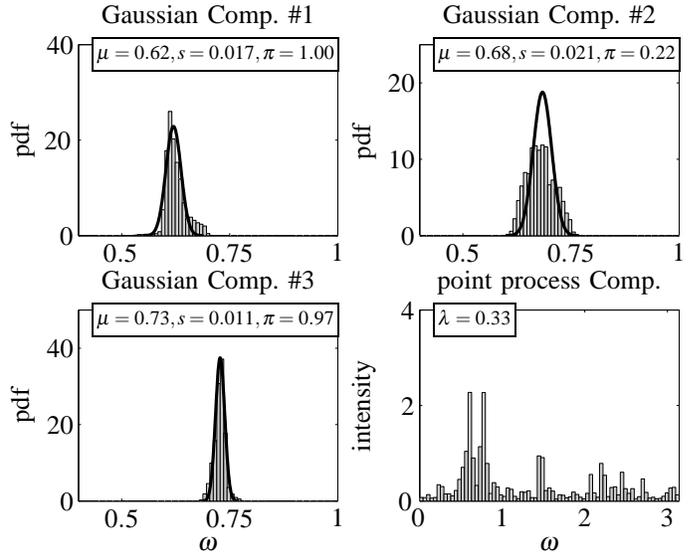
  \caption{Histogram of the labeled samples, that is, the samples
    allocated to the Gaussian and Poisson point process components,
    versus the pdf's of estimated Gaussian components in the model
    (black solid line) using \VAPoRS on the sinusoid detection
    example. The estimated parameters of each component are presented
    in the corresponding panel.}
  \label{fig:sin-alloc}
\end{figure}

As discussed in Section~\ref{sec:intro}, one of the main objectives of
the algorithm we proposed is to solve the label-switching issue in a
trans-dimensional setting. Figures~\ref{fig:sin-alloc} shows the
histograms of the labeled samples, i.e., $(\xxM{i},\,\zzM{i})$,
with~$i = 1,\ldots,M$, along with the pdf's of the estimated Gaussian
components (black solid line).  Moreover, the summaries provided by
VAPORS for each component are presented in its corresponding panel. We
used the average of the last 50 SEM iterations as parameter estimates,
as recommended in the SEM literature; see, for example,
\cite{celeux:1992:SEM, nielsen:2000:SEM}.  
 Comparing the distributions of the labeled samples with the ones of
 the posterior distributions of the sorted radial frequencies
 given~$k=3$ shown in Figure~\ref{fig:visu}, which are highly
 multimodal, reveals the capability of \VAPoRS in solving
 label-switching in a variable-dimensional setting.

 Looking at the bottom right panel of Figure~\ref{fig:sin-alloc}, the
 role of the point process component in capturing the outliers in the
 observed samples, which cannot be described by the Gaussian
 components, becomes clearer. Note that, without the point process
 component, these outliers would be allocated to the Gaussian
 components and would, consequently, induce a significant deterioration
 of the parameter estimates.

Table~\ref{table:param} presents the summaries provided using \VAPoRS
along with the ones obtained using the BMS approach. Contrary to the
BMS approach, \VAPoRS has enabled us to benefit from the information
of all probable models to give summaries about the middle harder to
detect component. Turning to the results of \VAPoRS, it can be seen
that the estimated means are compatible with the true radial
frequencies. Furthermore, the estimated probabilities of presence are
consistent with uncertainty of them in the variable-dimensional
posterior shown in Figure~\ref{fig:visu}. 

\begin{table}[htb!]
  \centering
  \begin{tabular}{|c|c|c|c|c|c|}
    \hline
    Comp.&$\mu$&$s$&$\pi$&$\mu_{BMS}$&$s_{BMS}$\\ 
    \hline        
    $1$&$0.62$&$0.017$&$1$&$0.62$&$0.016$\\
    \hline
    $2$&$0.68$&$0.021$&$0.22$&---&---\\
    \hline
    $3$&$0.73$&$0.011$&$0.97$&$0.73$&$0.012$\\
    \hline
  \end{tabular}
  \caption{Summaries of the variable-dimensional posterior
    distribution shown in Figure~\ref{fig:visu}; \VAPoRS vs. the BMS
    approach. }
  \label{table:param}
\end{table}

\begin{figure}[t!]
  \centering
  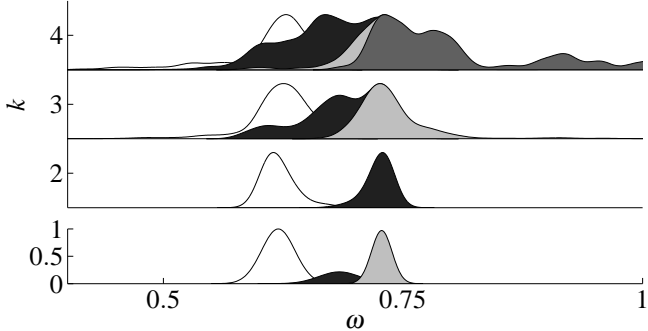
  \caption{Posterior distribution of the sorted radial
    frequencies~$\ok$ given~$k$ (top) and normalized pdf of the fitted
    Gaussian components (bottom).}
  \label{fig:sin-valid}
\end{figure}

To observe better the ``goodness-of-fit'' of the estimated Gaussian
components, the bottom panel of Figure~\ref{fig:sin-valid} depicts
their normalized densities\footnote{To obtain the normalized
  densities, first, we normalized the estimated pdf's to have their
  maximum equal to one. Then, we multiplied the estimated probability
  of presence of each Gaussian component to its corresponding
  normalized estimated pdf. Thus, the maximum of each normalized
  density is equal to the corresponding estimated probability of
  presence.}, under the posterior distributions of the sorted radial
frequencies given~$k$. This figure can be used to validate the
coherency of the estimated summaries with the information in the
variable-dimensional posterior distribution. It can be seen from the
figures that the shape of the pdf's of the estimated Gaussian
components are coherent in both the location and dispersion with the
ones of the posterior of the sorted radial frequencies.

\begin{figure}[t!]
  \centering
%
%
\begin{psfrags}%
\psfragscanon%
%
\psfrag{s03}[b][b]{\color[rgb]{0,0,0}\setlength{\tabcolsep}{0pt}\begin{tabular}{c}intensity\end{tabular}}%
\psfrag{s04}[t][t]{\color[rgb]{0,0,0}\setlength{\tabcolsep}{0pt}\begin{tabular}{c}$\omega$\end{tabular}}%
%
\psfrag{x01}[t][t]{$0.4$}%
\psfrag{x02}[t][t]{$0.6$}%
\psfrag{x03}[t][t]{$0.8$}%
\psfrag{x04}[t][t]{$1$}%
%
\psfrag{v01}[r][r]{$0$}%
\psfrag{v02}[r][r]{$10$}%
\psfrag{v03}[r][r]{$20$}%
\psfrag{v04}[r][r]{$30$}%
\psfrag{v05}[r][r]{$40$}%
%
\includegraphics[width=7cm]{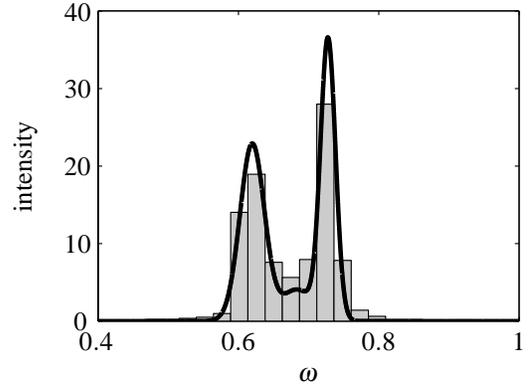}%
\end{psfrags}%
%

  \caption{Histogram intensity of all radial frequencies samples using
    the BMA approach along with the intensity of the fitted parametric
    model obtained using \VAPoRS.}
  \label{fig:sin-BMA}
\end{figure}

It is also useful for validating the estimated summaries to compare
the intensity of the estimated parametric model~$q_{\E}$ defined, in
general, as
\begin{equation}\label{eq:tap-intensity}
  h(\E) \;=\; \sum_{l=1}^{L} \pi_l \,\cdot\, \distrGauss(\dotvar \,|\,\m_l, \s_l),
\end{equation}
where we ignore the point process component, with the histogram
intensity of all radial frequencies obtained using the BMA approach
(see~\cite[Chapter~2]{roodaki:2012:phd} for more
information). Figure~\ref{fig:sin-BMA} shows such a figure for the
specific example of this section where the solid black line indicates
the intensity of the estimated parametric model. These figures also
indicate the ``goodness-of-fit'' of the fitted approximate posterior
and the true one.

\begin{figure}[t!]
  \centering 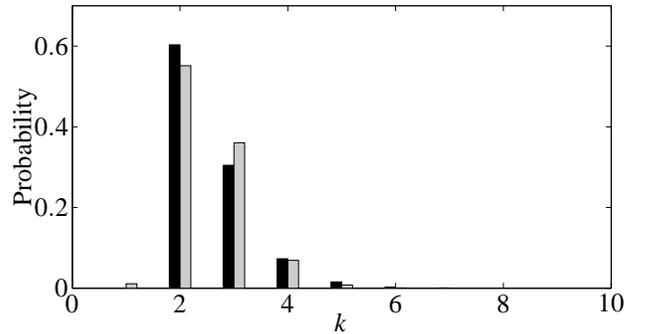
  \caption{Posterior distribution of the number~$k$ of number of
    components (black) and its approximated version (gray) obtained
    from the fitted model.}
  \label{fig:sin-post}
\end{figure}

Finally, to validate both the estimated probabilities of presence of
the Gaussian components and the mean parameter~$\lambda$ of the
Poisson point process component, Figure~\ref{fig:sin-post} illustrates
the posterior distribution of the number~$k$ of components together
with its approximated versions using \VAPoRS.  It can be seen from the
figure that \VAPoRS well captured the information provided in the true
posterior of the number~$k$ of components.

\subsection{Joint Bayesian detection and estimation of astrophysical
  particles in the Auger project}\label{sec:result-auger}

As the second illustrative example, we show results on a signal
decomposition problem encountered in the international astrophysics
collaboration called Auger \cite{auger:1997:report,
  albrowd:2004:auger}. The Auger project is aimed at studying
ultra-high energy cosmic rays, with energies in order of $10^{19}$eV,
the most energetic particles found so far in the universe. The
long-term objective of this project is to study the nature of those
ultra-high energy particles and determine their origin in the
universe. Nevertheless, they are \emph{not} observed directly. In
fact, when they collide the earth's atmosphere, a host of secondary
particles are generated, some of which, mostly ``muons'', finally
reach the ground.  To detect them, the Pierre Auger Cosmic Ray
Observatory was built which consists of two independent detectors; an
array of Surface Detectors (SD) and a number of Fluorescence Detectors
(FD).

The number of muons and their arrival times can be used as indications
of both the chemical composition and the origin of the primary
particles (see~\cite{auger:1997:report, albrowd:2004:auger} for more
information). Here, we concentrate on the signal decomposition
problem, where the goal is to count the number of muons and estimate
their individual parameters from the signals observed by SD detectors.
To show results, we use the Bayesian algorithm and the RJ-MCMC sampler
developed in~\cite{balazs-2008-bayes, bard-2010-single,
  bard-2012-adapt} for the trans-dimensional problem of joint
detection and estimation of muons. In this section, we first briefly
describe the problem and then use \VAPoRS to relabel and summarize
variable-dimensional output samples of the RJ-MCMC sampler developed
by~\cite{balazs-2008-bayes, bard-2010-single, bard-2012-adapt}.

When a muon crosses a SD tank, it generates photoelectrons (PE's)
along its track that are, then, captured by detectors and create a
discrete observed signal.
We denote the vector of observed signal by~$\n= (n_1,\ldots,n_N) \in
\Nset^N$, where the element~$n_i$ indicates the number of PE's
deposited by the muons in the time interval
\begin{equation*}
  [t_{i-1},\,t_i) \;\triangleq\; [t_0 +(i-1)t_{\Delta},\,t_0+i\,t_{\Delta}),
\end{equation*}
where~$t_0$ is the absolute starting time of the signal
and~$t_{\Delta} = 25$~ns is the signal resolution (length of one bin).

\begin{figure}[t!]
  \centering 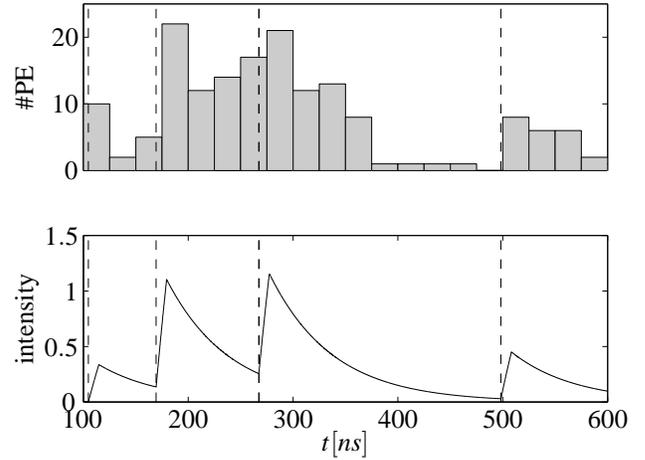
  \caption{(top) Observed signal~$\n$. (bottom) Intensity of the model
    $h(t \,|\, \am,\,\tm)$ defined
    in~\eqref{eq:auger-intensity}. There are~$k = 5$ muons in the
    signal with the true arrival times, i.e., $\tm = (105,\, 169,\,
    267,\, 268,\, 498)$, indicated by vertical dashed lines.}
  \label{fig:auger-obs}
\end{figure}

Each muon has two component-specific parameters, namely, the arrival
time~$t_{\mu}$ and the signal amplitude~$a_{\mu}$. The absorption
process of the photons generated by a muon is modeled by a
non-homogeneous Poisson point process with
intensity~\cite[Section~2.2]{bard-2010-single}
\begin{equation}\label{eq:auger-intensity}
  h(t \,|\, a_\mu, t_\mu) \;=\; a_\mu \, p_{\tau,t_d} (t - t_\mu),
\end{equation}
where~$ p_{\tau,t_d} (t)$ is the time response distribution,
$t_d$ is the rise-time and~$\tau$ is the exponential decay (both
measured in ns); see Figure~\ref{fig:auger-obs} (bottom) for such
exponential shape intensities.
Then, the expected number of PE's in the bin~$i$ is obtained by
integrating the intensity~\eqref{eq:auger-intensity} in the
corresponding bin:
\begin{equation}
  \bar{n}_i(a_{\mu},\,t_{\mu} ) 
  \;=\; a_{\mu} \int_{t_{i-1}}^{t_i}   p_{\tau,t_d}(t - t_{\mu})\mathrm{d}t.
\end{equation}

Conditioning on the number~$k$ of muons and the vector of
parameters~$\bm{t}_{\mu} = (t_{\mu,1}, \ldots, t_{\mu,k})$
and~$\bm{a}_{\mu} = (a_{\mu,1}, \ldots, a_{\mu,k})$, and assuming that
the number of PE's in each bin are independent, the likelihood is
written as
\begin{equation}\label{eq:auegr-like}
  p(\n \,|\, k, \bm{t}_{\mu}, \, \bm{a}_{\mu}) \;=\;
  \prod_{i=1}^N p(n_i\,|\,\bar{n}_i(k, \bm{a}_{\mu},\,\bm{t}_{\mu})), 
\end{equation}
where~$ p(n_i\,|\,\bar{n}_i(k, \bm{a}_{\mu},\,\bm{t}_{\mu} ))$ is a
Poisson distribution with the mean~$\bar{n}_i(k, \am,\,\tm)$. Then,
assuming independence of the muons, the expected number of PE's in
the~$i^\th$ bin, i.e., $\bar{n}_i(k, \am,\,\tm)$, given~$k$,
$\tm$, and $\am$ becomes
\begin{equation}
  \bar{n}_i(k, \am,\,\tm ) \;=\; \sum_{j=1}^k \bar{n}_i(a_{\mu,j},\,t_{\mu,j}).
\end{equation}

\begin{figure}[t!]
  \centering
  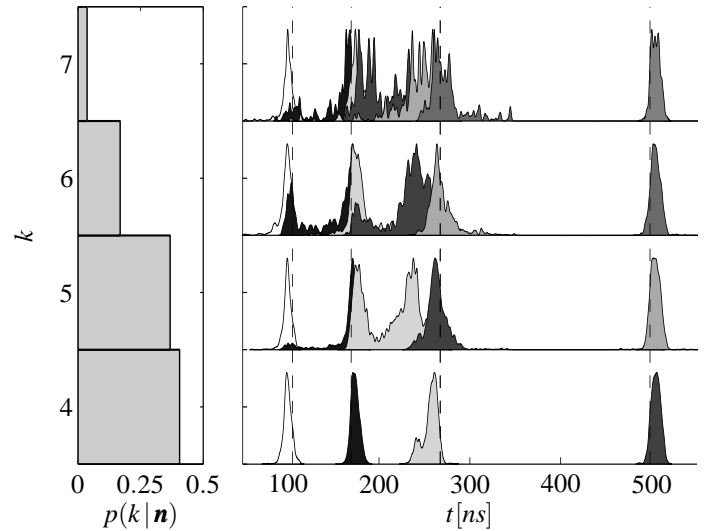
  \caption{Posterior distributions of the number $k$ of muons (left)
    and the sorted arrival times, $\tm$, given $k$ (right) constructed
    using 60\,000 RJ-MCMC output samples after discarding the burn-in
    period. The true number of components is five. The vertical dashed
    lines in the right figure locate the arrival times.}
  \label{fig:auger-visu}
\end{figure}

We will now illustrate the performance of \VAPoRS on a simulated PE
counting signal (see~\cite[Chapter~4]{roodaki:2012:phd} for results on
two other simulated experiments).  The observed signal of the
illustrative example considered here consists of five muons located at
$\tm = (105,\, 169,\, 267,\, 268,\, 498)$ (see
Figure~\ref{fig:auger-obs}). The posterior distributions of the
number~$k$ of muons and sorted arrival times are shown in
Figure~\ref{fig:auger-visu}. Note that, in this example, there are two
muons with almost equal arrival times, i.e., the third and fourth
muons.

Using the BMS approach, the model with four muons would be selected
($p(k = 4 \,|\, \n) = 0.4$), although~$\model_5$ has an almost similar
posterior probability of~0.38. Moreover, observe that the marginal
posterior of the arrival time of the third component is bimodal under
both~$\model_4$ and, more significantly so, $\model_5$.  We ran
\VAPoRS with~$L = 6$ Gaussian components
on the RJ-MCMC output samples shown in Figure~\ref{fig:auger-visu}
(note that~$p(k\leq 6 \,|\, \n) = 0.94$).

\begin{figure}[t!]
  \centering
  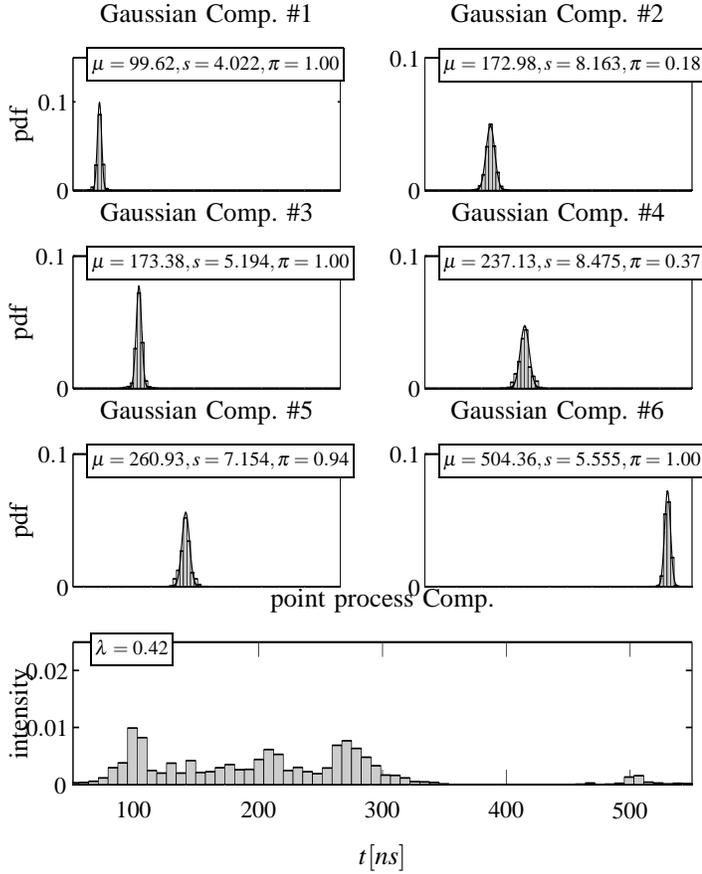
  \caption{Histogram of the labeled samples along with the pdf's of
    estimated Gaussian components in the model (black solid line)
    using \VAPoRS with~$L = 6$ on the variable-dimensional postrior
    shown in Figure~\ref{fig:auger-visu}. The estimated parameters of
    each component are presented in the corresponding panel.}
  \label{fig:auger-alloc}
\end{figure}

Figure~\ref{fig:auger-alloc} shows the histogram of the labeled
samples and the estimated parameters of the components. From the
figure, it can be seen that the bimodality effects caused by
label-switching exhibited in Figure~\ref{fig:auger-visu} is removed
completely and the estimated Gaussian components enjoy reasonable
variances. In the presented summary, there are four muons with high
probabilities of presence corresponding to the ones shown in the
bottom row of Figure~\ref{fig:auger-visu}. There are also two other
muons with comparatively low probabilities of presence.

In fact, the samples allocated to the point process component shown
the bottom row of Figure~\ref{fig:auger-alloc} can be regarded as the
residuals of the fitted model, that is, the observed samples which
the~$L$ Gaussian components in~$q_{\E}$ have not been able to
describe. These residuals can be used, as usual in statistics, as a
tool for goodness-of-fit diagnostics and model choice.

\begin{figure}[t!]
  \centering
%
%
\begin{psfrags}%
\psfragscanon%
\newcommand{\Xtick}[2]{\psfrag{#1}[t][t]{\footnotesize #2}}
\newcommand{\Ytick}[2]{\psfrag{#1}[r][r]{\footnotesize #2}}
\psfrag{s07}[l][l][0.75]{\setlength{\fboxsep}{2pt}\fcolorbox[rgb]{0,0,0}{1,1,1}{\color[rgb]{0,0,0}\setlength{\tabcolsep}{0pt}\begin{tabular}{l}$\lambda = 1.91$\end{tabular}}}%
\psfrag{s08}[b][b][0.75]{\color[rgb]{0,0,0}\setlength{\tabcolsep}{0pt}\begin{tabular}{c}intensity\end{tabular}}%
\psfrag{s09}[b][b][0.75]{\color[rgb]{0,0,0}\setlength{\tabcolsep}{0pt}\begin{tabular}{c}$L = 3$\end{tabular}}%
\psfrag{s10}[l][l][0.75]{\setlength{\fboxsep}{2pt}\fcolorbox[rgb]{0,0,0}{1,1,1}{\color[rgb]{0,0,0}\setlength{\tabcolsep}{0pt}\begin{tabular}{l}$\lambda = 0.92$\end{tabular}}}%
\psfrag{s11}[b][b][0.75]{\color[rgb]{0,0,0}\setlength{\tabcolsep}{0pt}\begin{tabular}{c}\end{tabular}}%
\psfrag{s12}[b][b][0.75]{\color[rgb]{0,0,0}\setlength{\tabcolsep}{0pt}\begin{tabular}{c}$L = 4$\end{tabular}}%
\psfrag{s13}[l][l][0.75]{\setlength{\fboxsep}{2pt}\fcolorbox[rgb]{0,0,0}{1,1,1}{\color[rgb]{0,0,0}\setlength{\tabcolsep}{0pt}\begin{tabular}{l}$\lambda = 0.42$\end{tabular}}}%
\psfrag{s14}[b][b][0.75]{\color[rgb]{0,0,0}\setlength{\tabcolsep}{0pt}\begin{tabular}{c}intensity\end{tabular}}%
\psfrag{s15}[t][t][0.75]{\color[rgb]{0,0,0}\setlength{\tabcolsep}{0pt}\begin{tabular}{c}$t [ns]$\end{tabular}}%
\psfrag{s16}[b][b][0.75]{\color[rgb]{0,0,0}\setlength{\tabcolsep}{0pt}\begin{tabular}{c}$L = 6$\end{tabular}}%
\psfrag{s17}[l][l][0.75]{\setlength{\fboxsep}{2pt}\fcolorbox[rgb]{0,0,0}{1,1,1}{\color[rgb]{0,0,0}\setlength{\tabcolsep}{0pt}\begin{tabular}{l}$\lambda = 0.24$\end{tabular}}}%
\psfrag{s18}[b][b][0.75]{\color[rgb]{0,0,0}\setlength{\tabcolsep}{0pt}\begin{tabular}{c}\end{tabular}}%
\psfrag{s19}[t][t][0.75]{\color[rgb]{0,0,0}\setlength{\tabcolsep}{0pt}\begin{tabular}{c}$t [ns]$\end{tabular}}%
\psfrag{s20}[b][b][0.75]{\color[rgb]{0,0,0}\setlength{\tabcolsep}{0pt}\begin{tabular}{c}$L = 8$\end{tabular}}%
%
\Xtick{x01}{$100$}%
\Xtick{x02}{$200$}%
\Xtick{x03}{$300$}%
\Xtick{x04}{$400$}%
\Xtick{x05}{$500$}%
\Xtick{x06}{$100$}%
\Xtick{x07}{$200$}%
\Xtick{x08}{$300$}%
\Xtick{x09}{$400$}%
\Xtick{x10}{$500$}%
\Xtick{x11}{$0$}%
\Xtick{x12}{$0.1$}%
\Xtick{x13}{$0.2$}%
\Xtick{x14}{$0.3$}%
\Xtick{x15}{$0.4$}%
\Xtick{x16}{$0.5$}%
\Xtick{x17}{$0.6$}%
\Xtick{x18}{$0.7$}%
\Xtick{x19}{$0.8$}%
\Xtick{x20}{$0.9$}%
\Xtick{x21}{$1$}%
\Xtick{x22}{$0$}%
\Xtick{x23}{$0.1$}%
\Xtick{x24}{$0.2$}%
\Xtick{x25}{$0.3$}%
\Xtick{x26}{$0.4$}%
\Xtick{x27}{$0.5$}%
\Xtick{x28}{$0.6$}%
\Xtick{x29}{$0.7$}%
\Xtick{x30}{$0.8$}%
\Xtick{x31}{$0.9$}%
\Xtick{x32}{$1$}%
%
\Ytick{v01}{$0$}%
\Ytick{v02}{$0.02$}%
\Ytick{v03}{$0.04$}%
\Ytick{v04}{$0$}%
\Ytick{v05}{$0.02$}%
\Ytick{v06}{$0.04$}%
\Ytick{v07}{$0$}%
\Ytick{v08}{$0.02$}%
\Ytick{v09}{$0.04$}%
\Ytick{v10}{$0$}%
\Ytick{v11}{$0.02$}%
\Ytick{v12}{$0.04$}%
%
\includegraphics[width=09cm]{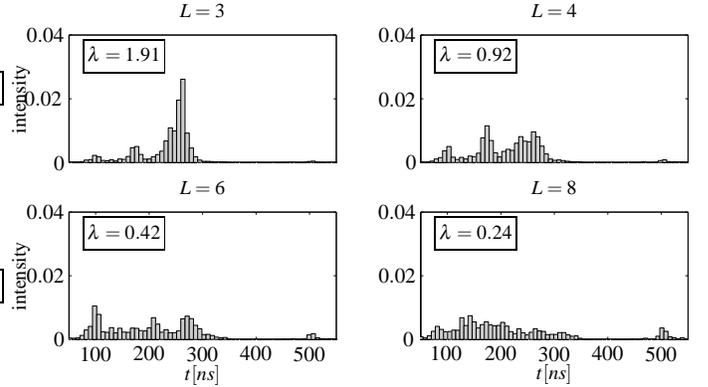}%
\end{psfrags}%
%

  \caption{Histograms of the residuals of the fitted model using
    \VAPoRS with different values of~$L = \{3, 4, 6, 8\}$.}
  \label{fig:auger-res}
\end{figure}

Figure~\ref{fig:auger-res} illustrates the histograms of the residuals
of the fitted model for different values of~$L \in \{3,\, 4,\, 6,\,
8\}$.  It can be seen from the top left panel of
Figure~\ref{fig:auger-res} that the distribution of the residuals
corresponding to the case where~$L = 3$ contains a few ``significant''
peaks. The peaks are gradually removed by adding Gaussian
components. When~$L = 4$, a component is added at~$t_\mu = 261$ that
captures samples distributed around the most significant peak of the
top left panel of Figure~\ref{fig:auger-res}. However, there still
exist a few peaks, particularly around~$t_\mu = 173$ which are
captured when~$L \geq 6$. However, the distribution of residuals for
the case of $L = 6$ and $L = 8$ do not differ significantly. Note the
decrease of value of $\hat{\lambda}$ by increasing $L$.

\begin{figure}
  \centering
  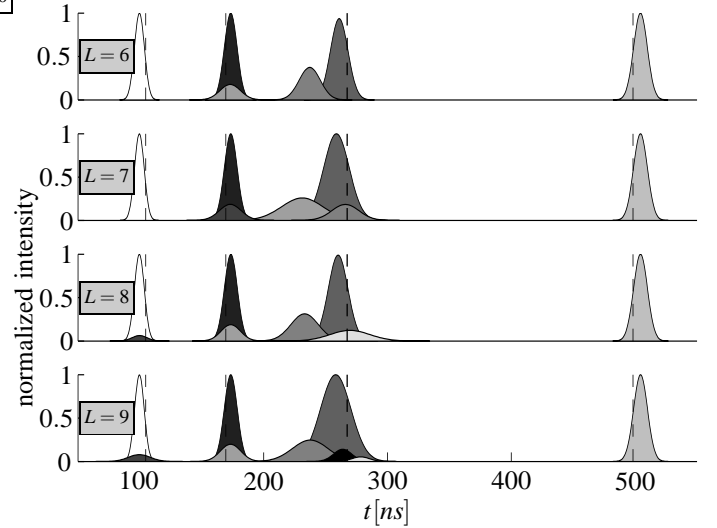
  \caption{Normalized pdf's of the fitted Gaussian components using
    \VAPoRS with different values of $6 \leq L \leq 9$.}
  \label{fig:auger-ve}
\end{figure}

Figure~\ref{fig:auger-ve} compares the normalized intensities of the
estimated Gaussian components for $6 \leq L \leq 9$. It can be seen
from the figure that changing~$L$ in a reasonable range, say, $6 \leq
L \leq 9$, does not influence significantly the final inference. In
all cases, the six Gaussian components that were estimated in the case
of~$L = 6$ exist. By moving from $L = 6$ to $L = 9$, additional
Gaussian components are added in the obtained summary with very low
probabilities of presence, which improve the fit but does not change
much the final inference.


\section{Monte Carlo experiment} \label{sec:result2}

The examples of Section~\ref{sec:result} have illustrated the
capability of \VAPoRS to relabel and summarize variable-dimensional
posterior distributions encountered in two signal decomposition
problems. In order to confirm these findings, we will now investigate
more systematically, by means of a Monte Carlo simulation experiment,
how faithfully the approximate posterior distribution preserves
certain features of the true posterior distribution.

One hundred realizations of the sinusoid detection experiment
described in Section~\ref{sec:example} (see Figure~\ref{fig:visu})
were simulated and analyzed using the same RJ-MCMC sampler as before.
The number of RJ-MCMC iterations was set to 100\,000 and the first
20\,000 samples were discarded as the burn-in period. Then, the
samples were thinned to one every fifth. To initialize the parametric
model~$q_{\E}$ in a systematic fashion, we set $L$ to the largest~$k$
such that its posterior probability is not less than 0.05. Then,
during the process of the SEM-type algorithms, if sufficient number of
samples, say, 10, is not allocated to a Gaussian component (or,
equivalently, its probability of presence fades to zero), we will
remove it from the parametric model and decrease~$L$ by one. Using
this approach generally results in approximate posterior distributions
which are ``richer'' than those provided by the BMS
approach\footnote{Later, in a post-processing step, since each
  Gaussian component has been endowed with a probability of
  presence~$\pi_l$, with~$1 \le l \le L$, one can decide to discard
  the ones with $\pi_l$ smaller than a certain threshold; see
  \cite[Section~3.4.3]{roodaki:2012:phd} for more discussion about
  this idea.}, in the sense that $L \ge \kMAP$, where $\kMAP =
\argmax_k \; p(k|\y)$. To initialize the Gaussian components'
parameters, i.e., the means $\mu_l$ and variances $s_l^2$, we used as
previously robust estimates of the mean and variances of the posterior
distributions of sorted radial frequencies given~$k=L$.

\begin{figure*}
  \centering %
  %
  %
  \begin{minipage}[b]{.5\linewidth} \centering
%
%
\begin{psfrags}%
\psfragscanon%
%
\psfrag{s03}[t][t]{\color[rgb]{0,0,0}\setlength{\tabcolsep}{0pt}\begin{tabular}{c}$\kMAP$\end{tabular}}%
\psfrag{s04}[b][b]{\color[rgb]{0,0,0}\setlength{\tabcolsep}{0pt}\begin{tabular}{c}$\kMAPhat$\end{tabular}}%
\psfrag{s05}[l][l]{\color[rgb]{0,0,0}\setlength{\tabcolsep}{0pt}\begin{tabular}{l}72\end{tabular}}%
\psfrag{s06}[l][l]{\color[rgb]{0,0,0}\setlength{\tabcolsep}{0pt}\begin{tabular}{l}2\end{tabular}}%
\psfrag{s07}[l][l]{\color[rgb]{0,0,0}\setlength{\tabcolsep}{0pt}\begin{tabular}{l}1\end{tabular}}%
\psfrag{s08}[l][l]{\color[rgb]{0,0,0}\setlength{\tabcolsep}{0pt}\begin{tabular}{l}24\end{tabular}}%
\psfrag{s09}[l][l]{\color[rgb]{0,0,0}\setlength{\tabcolsep}{0pt}\begin{tabular}{l}1\end{tabular}}%
%
\psfrag{x01}[t][t]{$1$}%
\psfrag{x02}[t][t]{$3$}%
\psfrag{x03}[t][t]{$5$}%
%
\psfrag{v01}[r][r]{$1$}%
\psfrag{v02}[r][r]{$3$}%
\psfrag{v03}[r][r]{$5$}%
%
\includegraphics[width=7cm]{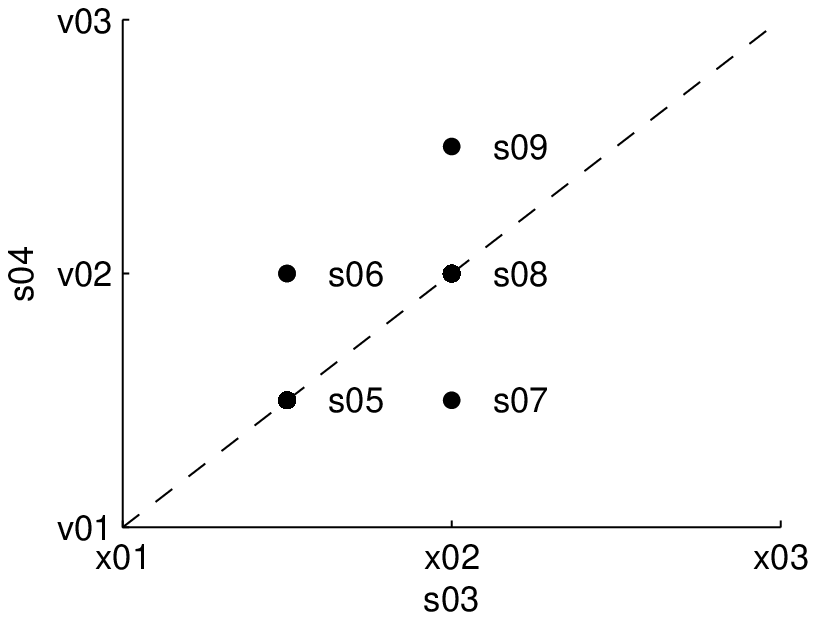}%
\end{psfrags}%
%
 \vspace{2mm}
    \subcaption{MAP estimator of~$k$}
  \end{minipage}%
  \begin{minipage}[b]{.5\linewidth} \centering
%
%
\begin{psfrags}%
\psfragscanon%
%
\psfrag{s03}[t][t]{\color[rgb]{0,0,0}\setlength{\tabcolsep}{0pt}\begin{tabular}{c}$p(k=2|\y)$\end{tabular}}%
\psfrag{s04}[b][b]{\color[rgb]{0,0,0}\setlength{\tabcolsep}{0pt}\begin{tabular}{c}$\hat{p}(k=2|\y)$\end{tabular}}%
%
\psfrag{x01}[t][t]{$0$}%
\psfrag{x02}[t][t]{$0.5$}%
\psfrag{x03}[t][t]{$1$}%
%
\psfrag{v01}[r][r]{$0$}%
\psfrag{v02}[r][r]{$0.5$}%
\psfrag{v03}[r][r]{$1$}%
%
\includegraphics[width=7cm]{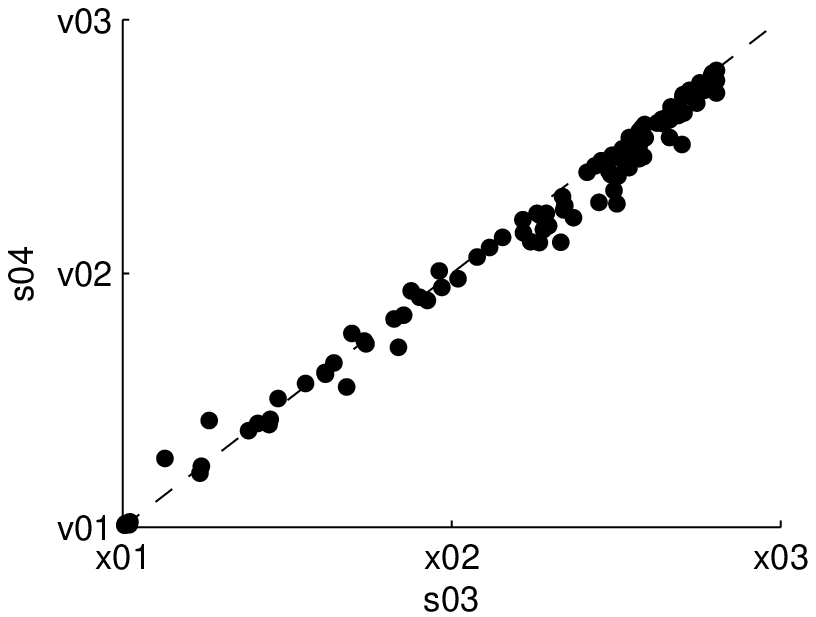}%
\end{psfrags}%
%
 \vspace{2mm}
    \subcaption{Posterior probability of~$\{ k = 2 \}$}
  \end{minipage}
  \vspace{3mm}
  %
  %
  \begin{minipage}[b]{.5\linewidth} \centering
%
%
\begin{psfrags}%
\psfragscanon%
%
\psfrag{s03}[t][t]{\color[rgb]{0,0,0}\setlength{\tabcolsep}{0pt}\begin{tabular}{c}$p(k=3|\y)$\end{tabular}}%
\psfrag{s04}[b][b]{\color[rgb]{0,0,0}\setlength{\tabcolsep}{0pt}\begin{tabular}{c}$\hat{p}(k=3|\y)$\end{tabular}}%
%
\psfrag{x01}[t][t]{$0$}%
\psfrag{x02}[t][t]{$0.5$}%
\psfrag{x03}[t][t]{$1$}%
%
\psfrag{v01}[r][r]{$0$}%
\psfrag{v02}[r][r]{$0.5$}%
\psfrag{v03}[r][r]{$1$}%
%
\includegraphics[width=7cm]{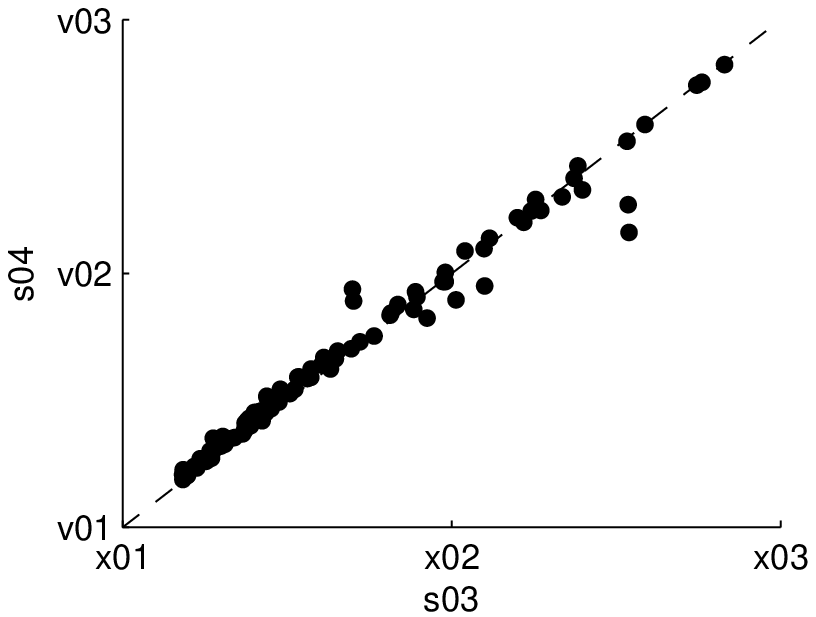}%
\end{psfrags}%
%
 \vspace{2mm}
    \subcaption{Posterior probability of~$\{ k = 3 \}$}
  \end{minipage}%
  \begin{minipage}[b]{.5\linewidth} \centering
%
%
\begin{psfrags}%
\psfragscanon%
%
\psfrag{s03}[t][t]{\color[rgb]{0,0,0}\setlength{\tabcolsep}{0pt}
  \begin{tabular}{c}direct BMA\end{tabular}}%
\psfrag{s04}[b][b]{\color[rgb]{0,0,0}\setlength{\tabcolsep}{0pt}
  \begin{tabular}{c}\VAPoRS\end{tabular}}%
%
\psfrag{x01}[t][t]{$-25$}%
\psfrag{x02}[t][t]{$-15$}%
\psfrag{x03}[t][t]{$-5$}%
%
\psfrag{v01}[r][r]{$-25$}%
\psfrag{v02}[r][r]{$-15$}%
\psfrag{v03}[r][r]{$-5$}%
%
\includegraphics[width=7cm]{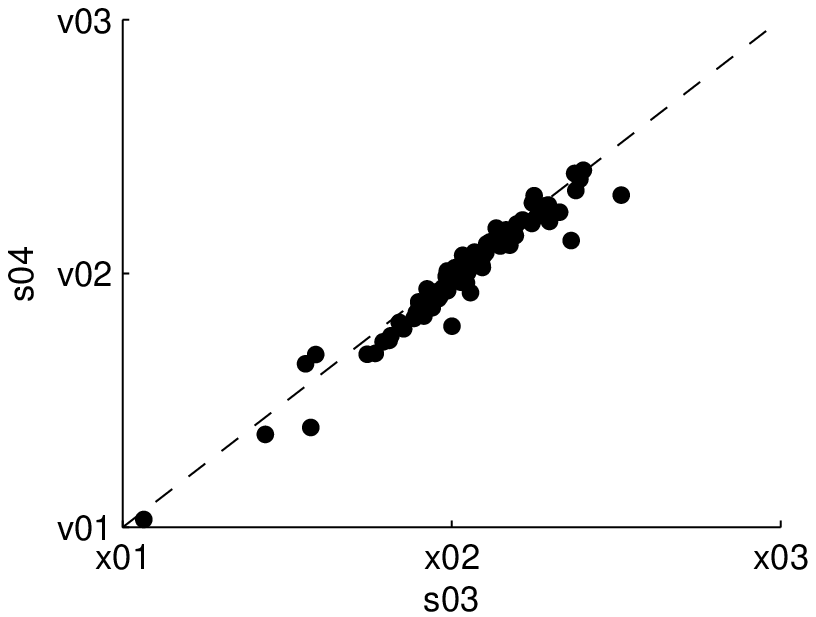}%
\end{psfrags}%
%
 \vspace{2mm}
    \subcaption{Reconstruction error (dB)}
  \end{minipage}
  \vspace{3mm}
  %
  %
  \begin{minipage}[b]{.5\linewidth} \centering
%
%
\begin{psfrags}%
\psfragscanon%
%
\psfrag{s03}[t][t]{\color[rgb]{0,0,0}\setlength{\tabcolsep}{0pt}
  \begin{tabular}{c}direct BMA\end{tabular}}%
\psfrag{s04}[b][b]{\color[rgb]{0,0,0}\setlength{\tabcolsep}{0pt}
  \begin{tabular}{c}\VAPoRS\end{tabular}}%
%
\psfrag{x01}[t][t]{$1$}%
\psfrag{x02}[t][t]{$2$}%
\psfrag{x03}[t][t]{$3$}%
\psfrag{x04}[t][t]{$4$}%
%
\psfrag{v01}[r][r]{$1$}%
\psfrag{v02}[r][r]{$2$}%
\psfrag{v03}[r][r]{$3$}%
\psfrag{v04}[r][r]{$4$}%
%
\includegraphics[width=7cm]{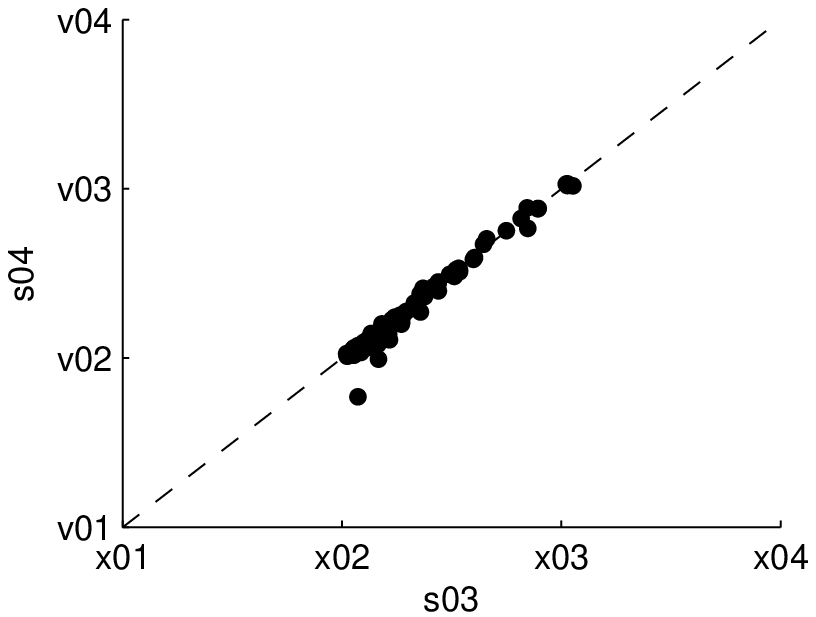}%
\end{psfrags}%
%
 \vspace{2mm}
    \subcaption{$\mathbb{E}(N(0, \pi/4))$}
  \end{minipage}%
  \begin{minipage}[b]{.5\linewidth} \centering
%
%
\begin{psfrags}%
\psfragscanon%
%
\psfrag{s03}[t][t]{\color[rgb]{0,0,0}\setlength{\tabcolsep}{0pt}
  \begin{tabular}{c}direct BMA\end{tabular}}%
\psfrag{s04}[b][b]{\color[rgb]{0,0,0}\setlength{\tabcolsep}{0pt}
  \begin{tabular}{c}\VAPoRS\end{tabular}}%
%
\psfrag{x01}[t][t]{$0$}%
\psfrag{x02}[t][t]{$0.5$}%
\psfrag{x03}[t][t]{$1$}%
%
\psfrag{v01}[r][r]{$0$}%
\psfrag{v02}[r][r]{$0.5$}%
\psfrag{v03}[r][r]{$1$}%
%
\includegraphics[width=7cm]{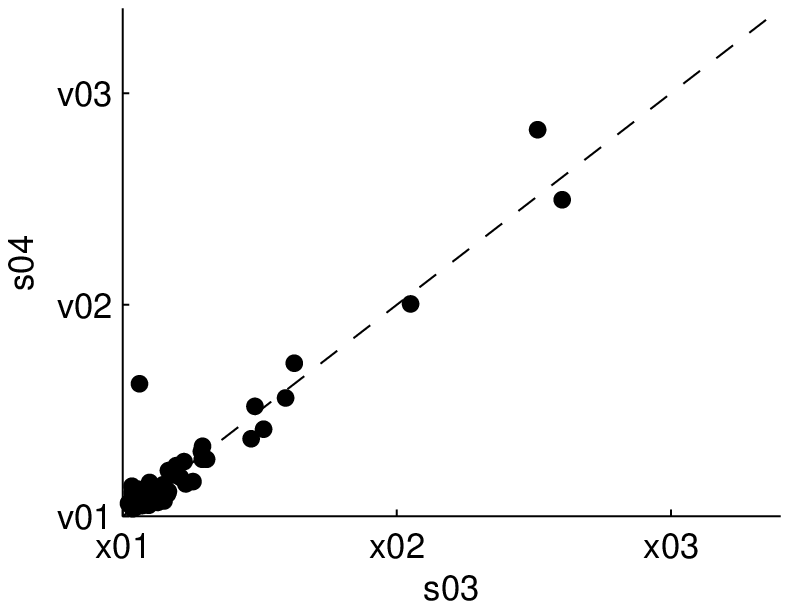}%
\end{psfrags}%
%
 \vspace{2mm}
    \subcaption{$\mathbb{E}(N(\pi/4, \pi/2))$}
  \end{minipage}
  \vspace{3mm}
  \caption{Comparison of (some features of) the true posterior
    distribution with its \VAPoRS approximation.}
  \label{fig:sin-compare-post}
\end{figure*}

Figure~\ref{fig:sin-compare-post} compares various features of the
fitted approximate posterior distribution~$q_{\hat{\E}}$, obtained
using 100 iterations of \VAPoRS, with the corresponding features of
the true variable-dimensional posterior distribution. These features
are described in the rest of this section.

The scatter plots shown in panels~(a), (b), and (c) compare the
posterior distribution of the number~$k$ of components, i.e.,
$p(k|\y)$, with its approximated version, denoted here by
$\hat{p}(k|\y)$, in 100 runs. We only show the posterior probabilities
of~$k=2$ and~$k=3$ in this comparison, as the other probabilities were
close to zero. The digits situated on the right of the points in the
panel~(a) indicate the number of occurrence of the corresponding event
in 100 runs and~$\kMAPhat = \argmax_k 
\hat{p}(k|\y)$. It can be seen from these three panels that the
information in~$p(k|\y)$ was well preserved by the approximated
posterior distributions.

Next we compare the performance of \VAPoRS with the one of the
``direct'' BMA approach\footnote{By ``direct'', we mean that posterior
  means are approximated using the RJ-MCMC samples directly, and not
  using the \VAPoRS posterior.} in reconstructing the noiseless
signal~$\y_0 = \D\; \ak$. To this end, the estimated reconstructed
noiseless signal is defined as
\begin{align}\label{eq:recons}
  \hat{\y}_0 & \;=\; \mathbb{E}(\y_0 \,|\, \y) \nonumber\\
  & \;=\; \sum_{k \in \Kcal} \int_{\THset^k}\, \mathbb{E}(\y_0 \,|\,
  k, \VectCompSpec, \y)\, p(k, \VectCompSpec \mid \y)\,
  \text{d}\VectCompSpec.
\end{align}
In the direct BMA approach, using the samples generated with the
RJ-MCMC sampler, the above integral is approximated by
\begin{equation*}
  \hat{\y}_0^{\BMA} \;=\; \frac{1}{M}\,
  \sum_{i=1}^M \D^{(i)}\; \hat{\bm{a}}_{1:k^{(i)}}^{(i)},
\end{equation*}
where~$\D^{(i)}$ is the design matrix of the~$i^\th$ vector of the
sampled radial frequencies~$\bm{\omega}_{1:k^{(i)}}^{(i)}$
and~$\hat{\bm{a}}_{1:k^{(i)}}^{(i)}$ is the posterior mean of the
amplitudes given $\bm{\omega}_{1:k^{(i)}}^{(i)}$ and its
hyperparameters. To reconstruct the noiseless signal from the fitted
approximate posterior~$q_{\hat{\E}}$ using \VAPoRS, one can
generate~$R$ pairs of samples~$(k^{(r)},
\bm{\omega}_{1:k^{(r)}}^{(r)})$ as explained in
Section~\ref{sec:model}. Then, we set
\begin{equation*}
  \hat{\y}_0^{\text{\VAPoRS}} = \frac{1}{R} \sum_{r=1}^R
  \D^{(r)}\; \hat{\bm{a}}_{1:k^{(r)}}^{(r)}.
\end{equation*}

Panel~(d) compares the normalized reconstruction errors when using
VAPORS with the ones of the direct BMA approach in dB, defined as
\begin{equation}
  10\,\log_{10}\left( \frac{\|\hat{\y}_0 - \y_0\|^2}{\|\y_0\|^2}\right),
\end{equation}
where~$\|\cdot\|$ is the L$_2$-norm and we set~$\hat{\y}_0 =
\hat{\y}_0^{\BMA}$ and~$\hat{\y}_0 = \hat{\y}_0^{\VAPoRS}$, when using
the BMA approach and \VAPoRS, respectively.  It can be seen from the
figure that the normalized errors of the reconstructed noiseless
signals using the compact summary obtained by \VAPoRS are quite
comparable with the ones obtained using the BMA approach.

Finally, the scatter plots in the last two panels compare the expected
number of components in the intervals~$(0,\pi/4)$ and~$(\pi/4,\pi/2)$
using \VAPoRS with, again, the ones obtained using the direct BMA
approach.  For the BMA approach, the expected number of components in
an interval~$T \subset \left( 0; \pi \right)$ is given by
\begin{equation*}
  \mathbb{E}(N(T) \mid \y)\;=\; \sum_{k\in \Kcal}
  \mathbb{E}(N(T) \mid k, \y) \, p(k \mid \y)
  \;\approx\; \frac{1}{M} \sum_{i = 1}^M N^{(i)}(T)\,,
\end{equation*}
where $N^{(i)}(T)$ is the number of radial frequencies observed in~$T$
on the~$i^\th$ sample.  On the other hand, from the summary
provided by \VAPoRS, the expected number of components in interval~$T$
is
\begin{equation*}
  \mathbb{E}_{\hat{\E}}\left(N(T) \mid \y\right) \;=\; 
  \sum_{l=1}^L \hat{\pi}_l\; \mathcal{N}(T; \hat{\E}_l)
  \;+\; \hat{\lambda}\, \frac{|T|}{|\THset|},
\end{equation*}
where~$\mathcal{N}(T; \hat{\E}_l)$ denotes the probability of~$T$
under the Gaussian distribution with parameters~$\hat{\E}_l$. The
figures confirm that the expected number of components in the chosen
intervals computed using both approaches are very similar.

The results shown in this section confirmed that the approximate
posterior distribution~$q_{\hat{\E}}$ obtained using \VAPoRS preserves
faithfully several important features of the true posterior
distribution; see~\cite[Section~3.4]{roodaki:2012:phd} for more
results in this vein, including a numerical investigation comparison
of the properties of estimators derived from \VAPoRS.

\section{Conclusion} \label{sec:conclusion} 

In this paper, we have proposed a novel algorithm to relabel and
summarize variable dimensional posterior distributions encountered in
signal decomposition problems when the number of component is
unknown. For this purpose, a variable-dimensional parametric model has
been designed to approximate the posterior of interest. The parameters
of the approximate model have been estimated by means of an SEM-type
algorithm, using samples from the true posterior distribution~$\post$
generated by a trans-dimensional Monte Carlo sampler, e.g., the
RJ-MCMC sampler. Modifications of our initial SEM-type algorithm have
been proposed, in order to cope with the lack of robustness of maximum
likelihood-type estimates.

The relevance of the proposed algorithm, both for summarizing and for
relabeling variable-dimensional posterior distributions, has been
illustrated on two signal decomposition examples, namely, the problem
of detection and estimation of sinusoids in Gaussian white noise and a
particle counting problem motivated by the astrophysics project
Auger. Most notably, \VAPoRS has been shown to be the first approach
in the literature capable of solving the label-switching issue in
trans-dimensional problems. We have shown that the proposed parametric
model provides a good approximation for the posteriors encountered in
both applications. Moreover, \VAPoRS can provide the user with more
insight concerning not only the component-specific parameters but also
the uncertainties about their presence.


We believe that this algorithm can be useful in the vast domain of
signal decomposition and mixture model analysis to enhance inference
in trans-dimensional problems. Theoretical investigations are required
in order to extend available existing convergence results for the SEM
algorithm to the SEM-type algorithm used in this paper (with
correlated input data and Metropolis-Hastings updates). Future work
will focus on using \VAPoRS to design more efficient adaptive
trans-dimensional MCMC methods, as a continuation of the ideas
presented in~\cite{bai:2011:divcon, bard-2012-adapt}.


\begin{center}
  \textsc{Acknowledgment}\\
\end{center}
The authors would like to express their gratitude to B. Kégl for his
collaboration and providing them with data for the Auger example.


\bibliographystyle{IEEEbib}
\scriptsize\setlength{\bibsep}{0.75ex}
\bibliography{REF}

\end{document}